\documentclass[twocolumn,prc,
showkeys,
 amsmath,amssymb,
 aps, 
]{revtex4-2}

\usepackage{graphicx}% Include figure files
\usepackage{dcolumn}% Align table columns on decimal point
\usepackage{bm}% bold math

\usepackage{mathrsfs}
\usepackage{amsmath}
\usepackage{color}
\usepackage{xcolor}
\usepackage{lineno}
\nolinenumbers
\DeclareMathOperator{\arctanh}{arctanh}

\usepackage{braket} 
\begin{document}

\preprint{APS/123-QED}

\title{\textbf{Collective-flow signatures of bubble nuclei in intermediate-energy projectile fragmentation reactions}
}% 

\author{Ge Ren}
\affiliation{Centre of Theoretical Physics, College of Physics, Henan Normal University, Xinxiang 453007, China}
\affiliation{Shanghai Advanced Research Institute, Chinese Academy of Sciences, Shanghai, 201210, China}
\author{Chun-Wang Ma}
\email[Corresponding author, ]{machunwang@126.com}
\affiliation{Institute of Nuclear Science and Technology, Henan Academy of Sciences, Zhengzhou 450046, China}
\affiliation{Centre of Theoretical Physics, College of Physics, Henan Normal University, Xinxiang 453007, China}
\affiliation{Shanghai Research Center for Theoretical Nuclear Physics, NSFC and Fudan University, Shanghai, 200438, China}
\author{Xi-Guang Cao}
\email[Corresponding author, ]{caoxg@sari.ac.cn}
\affiliation{Shanghai Advanced Research Institute, Chinese Academy of Sciences, Shanghai, 201210, China}
\affiliation{Shanghai Institute of Applied Physics, Chinese Academy of Sciences, Shanghai, 201800, China}
\affiliation{University of Chinese Academy of Sciences, Beijing 101408, China}
\author{Kai-Xuan Cheng}
\affiliation{Centre of Theoretical Physics, College of Physics, Henan Normal University, Xinxiang 453007, China}
\author{Jie Pu}
\affiliation{Centre of Theoretical Physics, College of Physics, Henan Normal University, Xinxiang 453007, China}

\date{\today}

\begin{abstract}

Characterized by central density depletion, bubble nuclei represent an important class of exotic nuclear structures. Collective flow in the $^{40}\mathrm{Ca}+^{9}\mathrm{Be}$ projectile fragmentation reaction at 140 MeV/u is investigated using the extended quantum molecular dynamics (EQMD) model with a parameter on of the isoscalar monopole excitation strength $\varepsilon$. The projectile $^{40}\mathrm{Ca}$ is initialized as a relaxed ten-$\alpha$ cluster configuration exhibiting a bubble-like density distribution. The results reveal that increasing $\varepsilon$ modifies the density distribution and the size of the depleted central region of bubble, producing pronounced changes in fragment yields and collective flow around $\varepsilon\approx0.5$. Below this value, heavy-fragment production plays a significant role, while the directed flow $v_1$ of protons and neutrons remains predominantly negative, indicating preferential emission toward the target side. Above $\varepsilon\approx0.5$, enhanced expansion of the central low-density region suppresses heavy residues and drives $v_1$ toward positive values. In contrast, preformed $\alpha$ clusters show different responses, reflecting their origin in the initial cluster configuration rather than solely in secondary decay. At low excitation, the cluster flow exhibits a pronounced impact-parameter dependence, whereas stronger excitation weakens cluster binding and suppresses directional anisotropy, driving the flow close to zero. The rapidity slopes of the flow coefficients also vary systematically with $\varepsilon$, tracing changes in emission anisotropy during expansion. These findings extend traditional static descriptions by probing the dynamical evolution of exotic structures in heavy-ion reactions. The contrasting collective-flow behavior of preformed clusters and single nucleons constitutes a sensitive dynamical probe of bubble-like structures, providing a complementary avenue for investigating exotic nuclear configurations in intermediate-energy heavy-ion collisions.

\end{abstract}

\maketitle

\section{Introduction}

The discovery of exotic nuclear structures has fundamentally reshaped our understanding of nuclear matter~\cite{Ye2025,cao2010PRC,cao2012PRC}. In the semi-classical liquid-drop model, atomic nuclei are typically depicted as nearly spherical systems with a uniform central density saturating at approximately $\rho_0 \approx 0.16~\mathrm{fm}^{-3}$. However, theoretical investigations indicate that, under specific conditions, such as high-spin excited states~\cite{B.Deyshan2020, Ichikawa2012}, antisymmetrization effects in light-mass nuclei~\cite{zhouboPRL2013, xufeilong2026}, and cluster configurations~\cite{yamaguchi2017, CaoXG2019, zhoubo2023NC, natowitz2026PRC}, nuclei can exhibit a variety of non-trivial geometric configurations. Among these intriguing phenomena, ``bubble nuclei'' have attracted significant theoretical and experimental attention. Characterized by a significant depletion of the central nucleon density, this concept was initially proposed by Wilson~\cite{Wilson1946,Wong1972}. Subsequently, extensive studies have explored its potential existence in medium-mass nuclei~\cite{MG2009, LiZP2014, VChoudhary2020} as well as in superheavy elements~\cite{Bender1999, Decharge2003}.

Previous studies have suggested that such bubble structures were predicted to emerge in nuclei with specific shell configurations, where the occupation of low-angular-momentum $s$ orbitals is suppressed~\cite{WangYZ2011, Saxena2019}. The interplay among shell effects, Coulomb repulsion, and nuclear interactions creates a delicate balance conducive to producing the hollow density distribution~\cite{VChoudhary2020, KHAN200837}. Some experiments have discussed whether Si isotopes exhibit a bubble structure, yet this issue remains under debate~\cite{AMutschler2017,ChenJ2024}.  However, a recent systematic study on the relaxed low-energy cluster configurations of all nuclides suggests that $^{40}\mathrm{Ca}$ and its neighboring isotopes with $A\approx40$ may serve as primary candidates for such bubble structures~\cite{RGnst2026,perera2022}. In this context, the formation of cluster states effectively inhibits the formation of a uniform spherical density profile, leading to central density depletion and making these nuclei ideal objects for probing the evolution of exotic structures in heavy-ion collisions~\cite{funaki2015, jin2026}. Consequently, identifying experimental observables capable of exploring these structural features in heavy-ion collisions is essential~\cite{Liao2024,cheng2024}.

Driven by developments in relevant theory and experiment, intermediate-energy projectile fragmentation reactions offer a robust dynamical probe of exotic nuclear structures~\cite{Ma21PPNP,xiaobaoNST2022,XuJS2026}, particularly the unique central density depletion of bubble nuclei. Unlike conventional compact nuclei with nearly saturated central densities, the initial ``hollow'' geometry of bubble nuclei may induce a distinct evolution of pressure gradients during the early compression stage, which is imprinted on the collective flow patterns of final-state particles. This dynamical response can be more sensitive to subtle structural differences during the transient high-density stage than static charge-radius measurements. To further amplify this effect and enhance observational sensitivity, the introduction of isoscalar monopole excitation as a modulation tool is crucial~\cite{RG2024PLB,Natowitz2024}. As an isotropic excitation mode, it enables systematic control over the size of the bubble-like region associated with central density depletion by modulating the nuclear matter distribution while preserving the overall nuclear symmetry. This process significantly magnifies the impact of initial structural discrepancies on collective-flow observables, providing a more discriminating experimental criterion for verifying the existence of bubble nuclei.

Extracting clear signatures of the bubble structure from fragmentation data requires a microscopic transport model capable of simultaneously describing the formation of exotic densities and the dynamical collision process. Standard coalescence models may lack the resolution to distinguish subtle density variations like a central bubble, while conventional mean-field approaches often struggle to capture the non-equilibrium dynamics of fragmentation. In this regard, the Extended Quantum Molecular Dynamics (EQMD) model provides a robust framework~\cite{HeWB2014, B.Deyshan2020}. By including variable Gaussian wave packet widths and a frictional cooling mechanism, EQMD can self-consistently generate the bubble-like structure and simulate its real-time evolution during the collision~\cite{RGnst2026}.

Previous works explored where such configurations emerge in the nuclear map and how they respond in isolation, but barely how they are encoded in reaction observables. In this work, the $^{40}\mathrm{Ca}+^{9}\mathrm{Be}$ projectile fragmentation reaction  at 140 MeV/u was simulated using the EQMD model to address this issue. With the fixed beam energy, the isoscalar monopole excitation strength $\varepsilon$ of the projectile is systematically varied to differ central depletion of nuclear matter distribution and the size of bubble, and track this effect through fragment production and the collective flow. This approach is designed to connect the intrinsic response of a bubble-like cluster configuration to its manifestation in reaction dynamics, and identify observables signaling the presence of central density depletion.

\section{Model and method}
\subsection{Extended quantum molecular dynamics model}
The quantum molecular dynamics (QMD) approach serves as a robust framework for investigating the dynamical evolution of nuclear reactions. While the conventional QMD model effectively describes properties of medium- to heavy-mass nuclei, it encounters limitations when applied to light nuclei because of its fixed Gaussian wave-packet width. To address these deficiencies, Maruyama \textit{et al.} introduced the extended QMD (EQMD) model in 1996~\cite{Maruyama}. The EQMD formalism incorporates critical improvements, including a frictional cooling mechanism for initialization, a variable wave-packet width parameter, and an explicit Pauli potential. These features enable a more reliable description of ground-state properties and cluster structures in light nuclei, which have been validated by heavy-ion collision experiments~\cite{HeWB2014,B.Deyshan2020}, making the framework particularly suitable for investigating exotic nuclear shapes, including bubble structures~\cite{RG2024PLB,RGnst2026}.

In EQMD, the total wave function of the $A$-nucleon system is approximated as a direct product of complex Gaussian wave packets,
\begin{equation}
  \Psi(\mathbf{r}_1, \dots, \mathbf{r}_A) = \prod_{i=1}^{A} \phi_i(\mathbf{r}_i),
  \label{eq:total_wavefunction}
\end{equation}
where each single-particle wave function $\phi_i$ is given by
\begin{equation}
  \phi_i(\mathbf{r}_i) = \left( \frac{\nu_i + \nu_i^*}{2\pi} \right)^{3/4} \exp\left[ -\frac{\nu_i}{2} (\mathbf{r}_i - \mathbf{R}_i)^2 + \frac{i}{\hbar} \mathbf{P}_i \cdot \mathbf{r}_i \right].
  \label{eq:single_wavepacket}
\end{equation}
Here, $\mathbf{R}_i$ and $\mathbf{P}_i$ denote the centroid position and momentum of the $i$-th nucleon, respectively. The complex width parameter $\nu_i$ is defined as
\begin{equation}
  \nu_i = \frac{1}{\lambda_i} + i\delta_i,
  \label{eq:width}
\end{equation}
where $\lambda_i$ and $\delta_i$ are real dynamical variables corresponding to the real and imaginary components, respectively; they govern the spatial spread and momentum-distribution width of the nucleon wave packet.

Based on the form of $\nu_i$, EQMD derives the time evolution of all $8A$ phase-space variables from the time-dependent variational principle,
\begin{equation}
\delta \int_{t_1}^{t_2} \langle \Psi | i\hbar \frac{\partial}{\partial t} - \hat{H} | \Psi \rangle \, dt = 0.
\label{eq:variational_principle}
\end{equation}
Applying this principle yields a set of classical Hamiltonian-like equations of motion for the centroid coordinates and momenta, as well as coupled equations for the width parameters~\cite{Maruyama},
\begin{align}
\dot{\mathbf{R}}_i &= \frac{\partial H}{\partial \mathbf{P}_i}, & \dot{\mathbf{P}}_i &= -\frac{\partial H}{\partial \mathbf{R}_i}, \label{eq:eom_rp} \\
\frac{3\hbar}{4} \dot{\lambda}_i &= -\frac{\partial H}{\partial \delta_i}, & \frac{3\hbar}{4} \dot{\delta}_i &= \frac{\partial H}{\partial \lambda_i}. \label{eq:eom_width}
\end{align}

A precise determination of the local nucleon density distribution $\rho(\mathbf{r})$ is fundamental to identifying exotic structural features, such as the central density depletion characteristic of bubble nuclei. In the EQMD framework, since the wave-packet width parameters $\lambda_i(t)$ and $\delta_i(t)$ evolve dynamically for each nucleon, the total density at any position $\mathbf{r}$ and time $t$ is constructed by summing the probability densities of all individual Gaussian wave packets with their instantaneous widths,
\begin{equation}
\rho(\mathbf{r}) = \sum_{i}^{A} \rho_i(\mathbf{r}),
\label{eq:density_sum}
\end{equation}
where the density contribution from the $i$-th nucleon with width parameter $\lambda_i$ and centroid $\mathbf{R}_i$ is given by
\begin{equation}
\rho_i(\mathbf{r}) = \frac{1}{(\pi \lambda_i)^{3/2}} \exp\left[ -(\mathbf{r} - \mathbf{R}_i)^2 / \lambda_i \right].
\label{eq:density_single}
\end{equation}

\subsection{Monopole excitation}
Unlike the experimental approach, in which isoscalar monopole excitations are induced through small-angle inelastic scattering with $\alpha$ particles, protons, $^{3}$He ions, EQMD employs a perturbative method to generate an excited configuration from the nuclear ground state~\cite{RG2024PLB, tao2013, yamada2012}. The total Hamiltonian of the system during excitation is decomposed into an unperturbed part $\hat{H}^{0}(t)$ and an applied excitation perturbation $\hat{H}'(t)$,

\begin{equation}
  \hat{H}(t) = \hat{H}^{0}(t) + \hat{H}'(t),
  \label{eq:total_hamiltonian}
\end{equation}
where the perturbation is defined as $\hat{H}'(t) = \varepsilon \hat{\mathcal{M}} \delta(t)$. Here, $\varepsilon$ determines the strength of the perturbation, and $\hat{\mathcal{M}}$ is the monopole excitation operator. Since the system contains $A$ nucleons, $\hat{\mathcal{M}}$ can be expressed as the sum of all single-particle operators $\hat{\mathscr{M}}_i$, and its expectation value is
\begin{equation}
  \langle \hat{\mathcal{M}} \rangle = \langle \Psi | \hat{\mathcal{M}} | \Psi \rangle = \sum_{i}^{A} \langle \Psi_i | \hat{\mathscr{M}}_i | \Psi_i \rangle.
  \label{eq:monopole_operator}
\end{equation}
The expectation value of $\hat{\mathcal{M}}$ can be calculated using the Wigner transform,
\begin{equation}
  \langle \hat{\mathcal{M}} \rangle = \iint f(\mathbf{r}, \mathbf{p}) \mathscr{M}(\mathbf{r}, \mathbf{p})
  \,d\mathbf{r} d\mathbf{p},
  \label{eq:wigner_transform}
\end{equation}
where $f(\mathbf{r}, \mathbf{p})$ is the Wigner distribution function and $\mathscr{M}(\mathbf{r}, \mathbf{p})$ is the classical counterpart of the operator $\hat{\mathscr{M}}$.

For the isoscalar monopole mode, the excitation corresponds to a scaling transformation in phase space. The monopole excitation operator is defined as $\hat{\mathcal{M}} = \sum_{i=1}^{A} r_i^2$, which characterizes the monopole transition strength. In the semiclassical QMD framework, the initial breathing motion is generated through a canonical scaling transformation in phase space. The corresponding phase-space generator can be written as the virial operator $\mathscr{M} = \mathbf{r}\cdot\mathbf{p}$, which produces a uniform radial expansion (or compression) together with the corresponding inverse scaling of the momentum distribution.

Applying the perturbation generated by $\mathscr{M} = \mathbf{r} \cdot \mathbf{p}$, the changes in coordinates and momenta are given by Hamilton's equations,
\begin{equation}
  \begin{cases}
      \mathbf{r}_i' = \mathbf{r}_i + \varepsilon \dfrac{\partial \mathscr{M}}{\partial \mathbf{p}_i}, \\[10pt]
      \mathbf{p}_i' = \mathbf{p}_i - \varepsilon \dfrac{\partial \mathscr{M}}{\partial \mathbf{r}_i}.
  \end{cases}
  \label{eq:scaling_derivation}
\end{equation}
The parameter $\varepsilon$ represents the excitation strength. This transformation uniformly scales the spatial distribution and inversely scales the momentum distribution, thereby generating a radially scaled initial configuration associated with the isoscalar monopole degree of freedom.

\subsection{Collective flow and kinematic observables}

The concept of collective flow in heavy-ion reactions was originally introduced in high-energy collisions of heavy nuclei \cite{ollitrault1992, reisdorf1997, ackermann2001}. Nevertheless, theoretical and experimental advances have verified that this phenomenon also manifests in intermediate-energy reactions \cite{xiaokui2023, weiSN2024, kireyeu2026} and in medium-mass \cite{zhangHY2002, yanTZ2006} and light-mass \cite{acharya2026, modak2026} nuclear systems. 

In the Quantum Molecular Dynamics (QMD) framework, the dynamical evolution of heavy-ion collisions is described by the time evolution of $N$ Gaussian wave packets. Macroscopic observables such as collective flow, transverse momentum, and rapidity are extracted from the phase-space distribution of nucleons at freeze-out or at selected simulation times \cite{WangJinMei2025}.

The transverse momentum $p_T$ relative to the beam axis ($z$ axis) for a particle $i$ is given by \cite{danielewicz1985}
\begin{equation}
  p_{T,i} = \sqrt{p_{x,i}^2 + p_{y,i}^2}.
\end{equation}

The azimuthal yield distribution of particles with respect to the reaction-plane angle $\Psi_R$ can be expanded in a Fourier series \cite{voloshin1996, voloshin2010},
\begin{equation}
  \frac{dN}{d\phi} \propto 1 + 2 \sum_{n=1}^{\infty} v_n \cos[n(\phi - \Psi_R)],
  \label{eq:fourier_flow}
\end{equation}
where the azimuthal angle is evaluated as $\phi = \operatorname{atan2}(p_y,p_x)$ for the particle momentum \cite{cox2024, lixin2026}. The coefficients $v_n$ represent the $n$-th order flow harmonics,

\vspace{0.5\baselineskip}

\indent
\hangafter=1
\hangindent=1em
\noindent
$v_1 = \langle \cos(\phi-\Psi_R) \rangle$: \textbf{Directed flow}, reflecting the sideward bounce-off of nuclear matter due to the initial geometric asymmetry in non-central collisions;

\indent
\hangafter=1
\hangindent=1em
\noindent
$v_2 = \langle \cos[2(\phi-\Psi_R)] \rangle$: \textbf{Elliptic flow}, indicating the conversion of the initial almond-shaped spatial deformation into momentum anisotropy driven by pressure gradients;

\indent
\hangafter=1
\hangindent=1em
\noindent
$v_3 = \langle \cos[3(\phi-\Psi_R)] \rangle$: \textbf{Triangular flow}, arising primarily from event-by-event fluctuations in the initial nucleon positions rather than the average collision geometry;

\indent
\hangafter=1
\hangindent=1em
\noindent
$v_4 = \langle \cos[4(\phi-\Psi_R)] \rangle$: \textbf{Quadrangular flow}, which contains contributions from both the initial fourth-order spatial eccentricity and nonlinear mode coupling.

\vspace{0.5\baselineskip}

The reaction-plane azimuthal angle, $\Psi_{\mathrm{RP}}$, is intrinsically determined by the direction of the impact-parameter vector $\mathbf{b}$. In experiments, where the impact-parameter direction is unknown event by event, event-plane or cumulant methods are required to estimate $\Psi_R$ and extract $v_n$ from the measured particle ensemble. However, in event-by-event aligned QMD simulations, it is a standard convention to align the collision axis along the $x$ direction (i.e., $\mathbf{b} = (b, 0, 0)$). Consequently, the reaction plane coincides with the $x$-$z$ plane, yielding a reaction-plane angle of $\Psi_{\mathrm{RP}} = 0$. The positive x-axis is defined toward the projectile side, and $v_1>0$ and $v_1<0$ denote preferential emission toward the projectile and target sides, respectively.

Rapidity $y$ is a kinematic variable preferred over velocity in relativistic heavy-ion collisions due to its additive property under Lorentz boosts along the beam direction. For a nucleon with energy $E_{i}$ and longitudinal momentum $p_{z,i}$, the rapidity is defined as
\begin{equation}
y_{i} = \frac{1}{2} \ln \left( \frac{E_{i} + p_{z,i}}{E_{i} - p_{z,i}} \right) = \arctanh \left( \frac{p_{z,i}}{E_{i}} \right),
\label{eq:rapidity}
\end{equation}
where the total energy $E_{i}$ is $\sqrt{m_{i}^{2} + p_{i}^{2}}$. In the non-relativistic limit applicable to intermediate energies, this reduces to the dimensionless longitudinal velocity, $y_i \approx p_{z,i}/m_i \simeq v_{z,i}/c$ when $c$ is restored.

\section{Results and discussion}

\begin{figure}
  \centering
  \includegraphics[width=\linewidth]{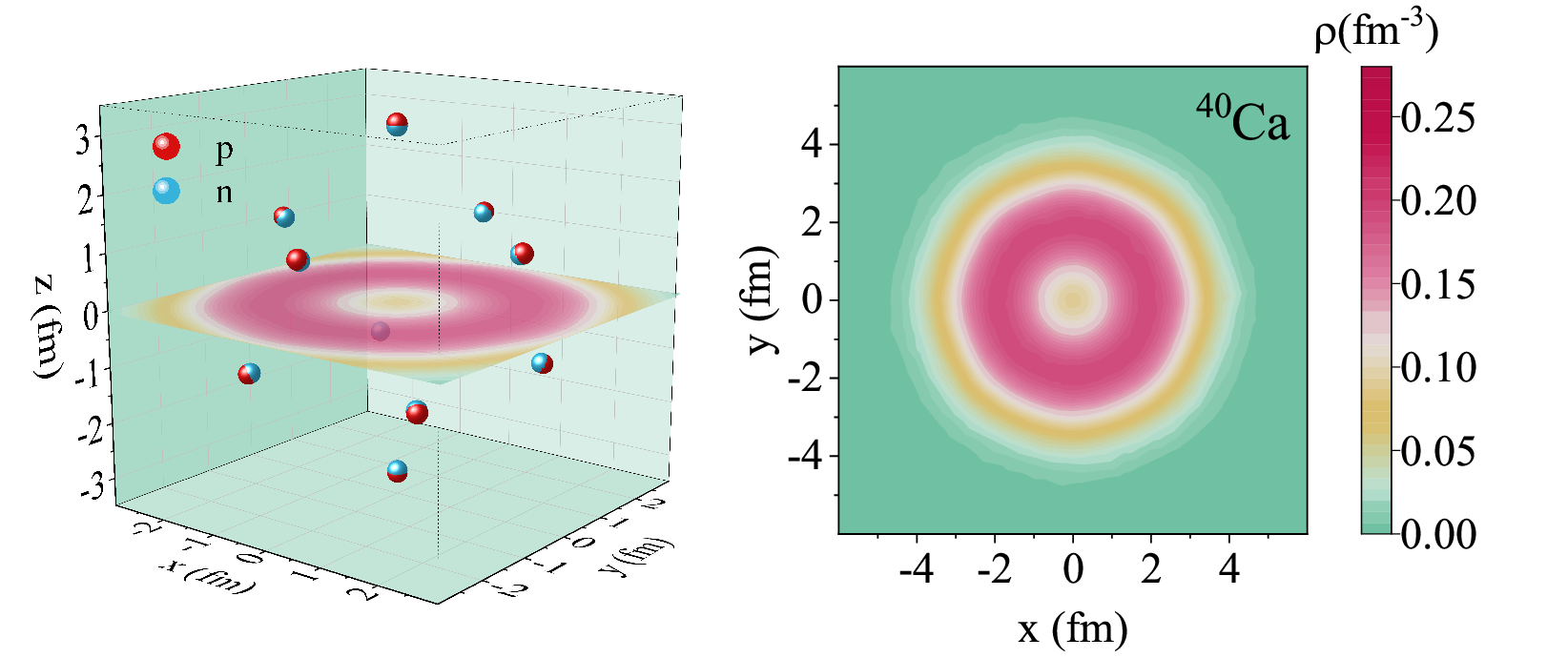}
  \caption{ Centroids of nucleon wave packets in coordinate space (left) for $^{40}\mathrm{Ca}$ in a relaxed low-energy cluster configuration with a bubble-like structure, together with the corresponding density distribution (right) at $z=0$.}
  \label{fig:40Ca}
\end{figure}

The collective flow was studied under deferent isospin monopole excitation strengths via EQMD simulations of the $^{40}\mathrm{Ca}+^{9}\mathrm{Be}$ projectile fragmentation reaction at 140~MeV/u. The $^{40}\mathrm{Ca}$ obtained from the EQMD initialization procedure corresponds to a relaxed low-energy ten-$\alpha$ cluster configuration exhibiting a bubble-like density distribution, as shown in Fig.~\ref{fig:40Ca}. Departing from the smooth density profiles of conventional mean-field approaches, inherent clustering triggers local density fluctuations and low-density inter-cluster gaps~\cite{Ebran2012, Zhoubo.prc.N13.2025, xufeilong2025}, which in turn drive significant deviations in collective-flow patterns \cite{maYG2023, guoCC2017, zhangS2017}. In this study, the monopole excitation strength $\varepsilon$ is employed to modulate the nuclear matter distribution and control the size of the bubble-like region in order to explore the effect of different degrees of central density depletion on collective flow.

\begin{figure}
  \centering
  \includegraphics[width=\linewidth]{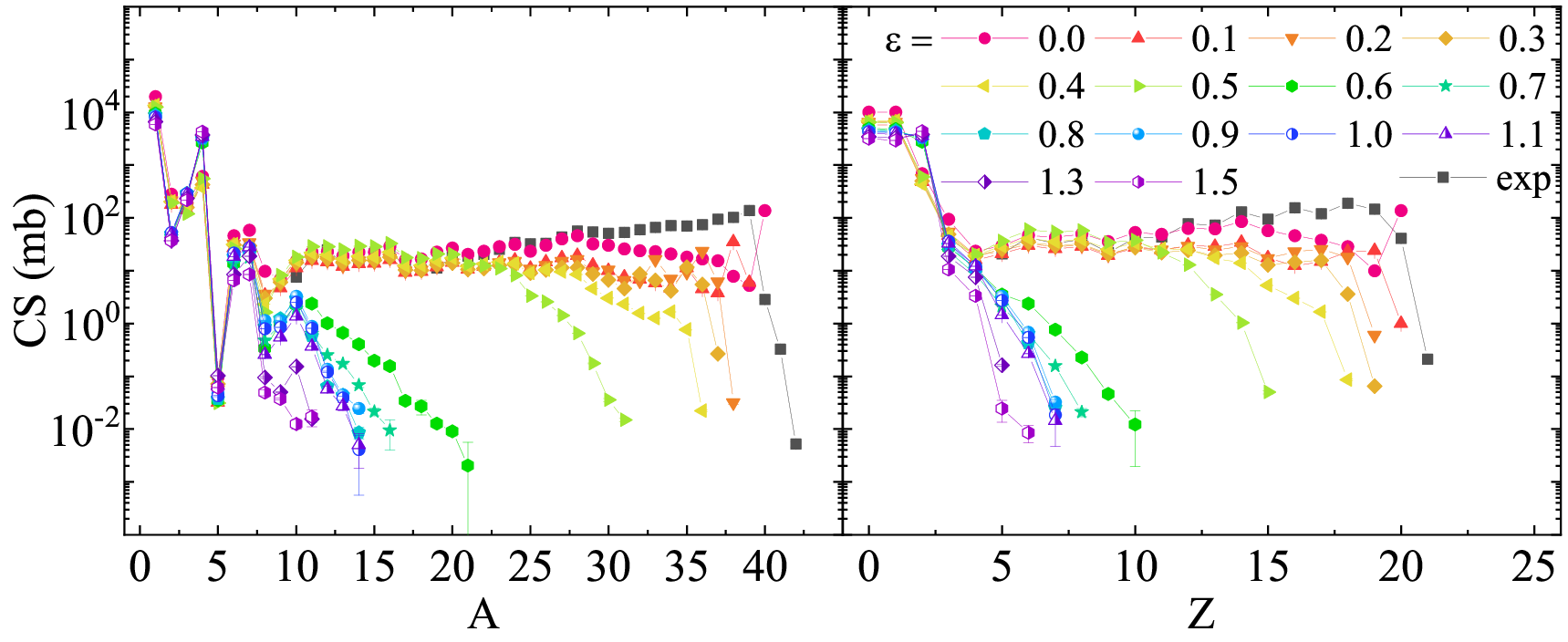}
  \caption{Production cross sections of isobars and isotopes in reactions induced by $^{40}\mathrm{Ca}$ under different excitation strengths $\varepsilon$. The black squares represent experimental data reported in Ref.~\cite{mocko2006}.}
  \label{fig:CS}
\end{figure}

Among many experimental observables, production cross sections provide a direct and experimentally accessible probe. Figures~\ref{fig:CS}~(a) and (b) present the production cross sections of isobars (with the same mass number $A$) and isotopes (with the same atomic number $Z$), respectively, produced in the reaction induced by $^{40}\mathrm{Ca}$ after coupling to the GEMINI statistical decay model under different excitation strengths with each $\varepsilon$ calculated 20000 events. The black squares represent the experimental data reported in Ref.~\cite{mocko2006}. The figures demonstrate a systematic evolution in which the cross sections for heavy fragments decrease gradually and eventually disappear as $\varepsilon$ increases, and a noticeable change in this trend appears around $\varepsilon\approx0.5$, consistent with the evolution of the density distribution reported in Ref.~\cite{RG2024PLB}. This behavior is associated with the fact that monopole excitation functions as a radial mode. Increased excitation strength enhances radial expansion of the projectile nucleus, which is accompanied by an expansion of the low-density central region and a reduction in the average binding of the system. As a result, the formation of heavy residual nuclei becomes unfavorable following projectile fragmentation, leading to an increased probability of producing light fragments relative to heavy residues. $^{40}\mathrm{Ca}$ with a relaxed low-energy cluster configuration at $\varepsilon=0.0$ already exhibits central-density depletion; therefore, the $\varepsilon=0.0$ calculation also follows this trend when compared with the experimental data. The enhancement of the cross sections at $A=40$ and $Z=20$ for the magenta circles may be associated with the enhanced survival probability of the doubly magic $^{40}\mathrm{Ca}$ during the statistical decay process in GEMINI.

\begin{figure}
  \centering
  \includegraphics[width=\linewidth]{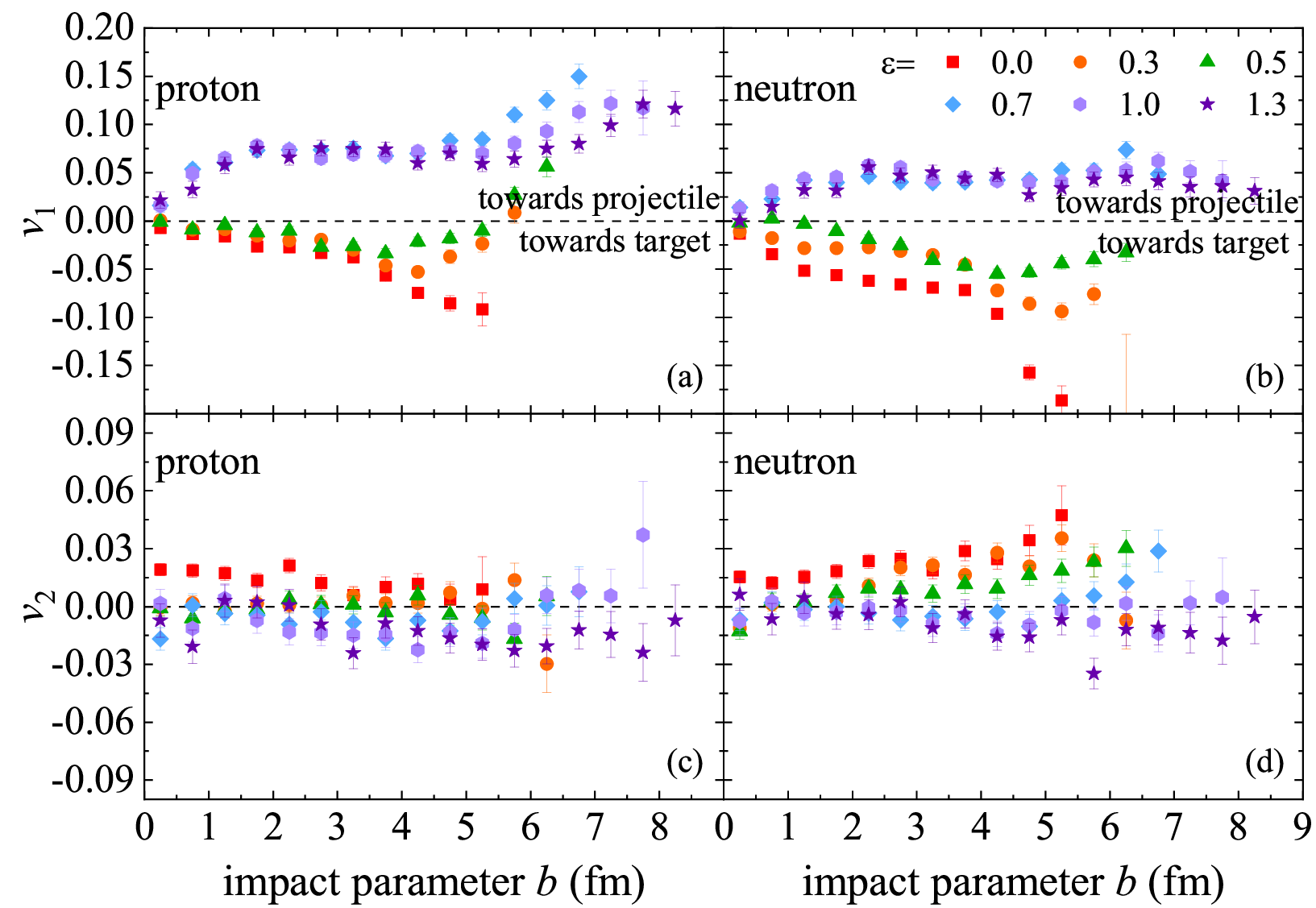}
  \caption{Directed $v_1$ and elliptic $v_2$ flows of protons and neutrons as functions of impact parameter $b$ for different isoscalar monopole excitation strengths $\varepsilon$. Increasing $\varepsilon$ is accompanied by a gradual evolution of $v_1$ from target- to projectile-directed, signaling enhanced radial expansion. Slightly negative $v_2$ at large $\varepsilon$ is consistent with a weak out-of-plane preference.
  }
  \label{fig:1H_1n_v12_b}
\end{figure}

Figure~\ref{fig:1H_1n_v12_b} presents the impact-parameter dependence of the directed flow $v_1$ and elliptic flow $v_2$ for protons (left panels) and neutrons (right panels) under different isoscalar monopole excitation strengths. In this figure, $v_1$ and $v_2$ exhibit a clear transition from negative values (toward the target side) at low $\varepsilon$ to positive values (toward the projectile side) at high $\varepsilon$, particularly for mid-to-large impact parameters. Notably, $\varepsilon = 0.5$ corresponds to the crossover region where $v_1$ crosses zero, which is consistent with our earlier observation that this value separates regimes dominated by residual-nucleus formation from those dominated by light-fragment production. For $v_1$ in panels (a) and (b), this reversal is consistent with the increasing radial expansion of the projectile nucleus with higher $\varepsilon$, which is associated with enhanced emission of light particles in the projectile direction, accompanied by weakened binding and enlargement of the central bubble. Panels (c) and (d) show the elliptic flow $v_2$, which remains close to zero with small fluctuations for most $b$ and $\varepsilon$ values, indicating only weak azimuthal anisotropy. However, for large $\varepsilon$, a slightly negative $v_2$ emerges, suggesting a weak out-of-plane emission preference akin to the squeeze-out pattern observed in heavy-ion collisions. This implies that the emission remains dominated by dynamical expansion rather than complete thermalization. Weaker binding may enhance surface emission, where particles escape preferentially out of the reaction plane from the liquid film region of bubbles during early-stage expansion. The out-of-plane preference for light particles may reflect their weaker Coulomb interaction and greater sensitivity to nuclear mean-field dynamics.

\begin{figure}
  \centering
  \includegraphics[width=\linewidth]{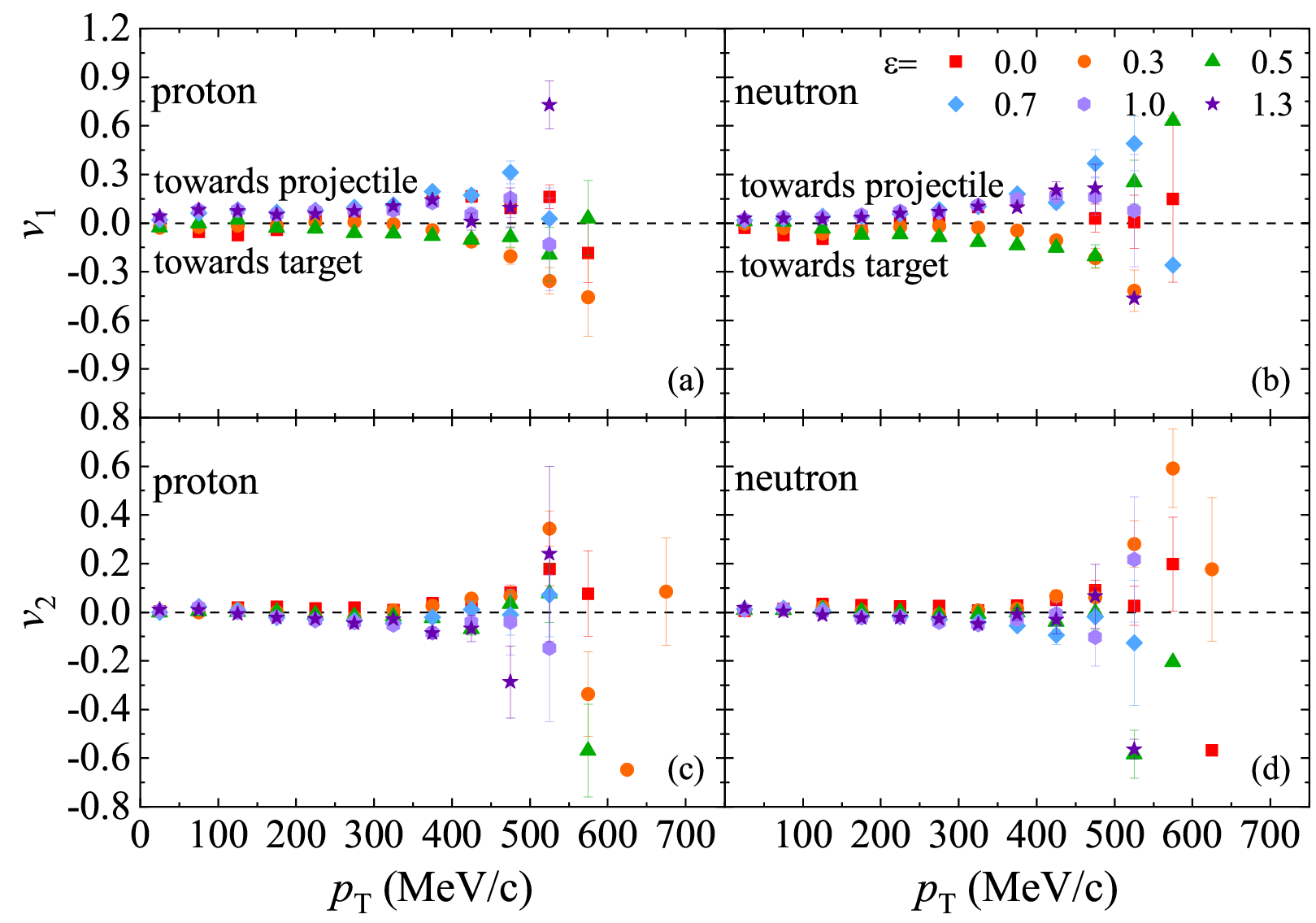}
  \caption{Directed flow $v_1$ and elliptic flow $v_2$ as functions of transverse momentum $p_T$ for protons (left) and neutrons (right) under different isoscalar monopole excitation strengths $\varepsilon$. The changes around $\varepsilon = 0.5$ are most pronounced at low $p_T$, while larger uncertainties at high $p_T$ for low $\varepsilon$ result from limited statistics due to suppressed high-momentum emission in weakly excited systems.
}
  \label{fig:1H_1n_v12_pT}
\end{figure}

The dependence of $v_1$ and $v_2$ on transverse momentum $p_T$, as shown in Fig.~\ref{fig:1H_1n_v12_pT}, follows patterns similar to those observed in Fig.~\ref{fig:1H_1n_v12_b}, with a change occurring around $\varepsilon = 0.5$. For values of $\varepsilon$ below 0.5, the directed flow is predominantly negative, indicating emission toward the target side, whereas for $\varepsilon$ above 0.5 it becomes positive, reflecting preferential emission toward the projectile side. This behavior is most evident in the low-$p_T$ region. At higher transverse momenta, the same sign-reversal trend generally persists, but data points corresponding to low excitation strengths exhibit larger uncertainties and greater scatter because fewer particles populate the high-$p_T$ region. This reduced high-$p_T$ yield is consistent with weaker radial excitation at low $\varepsilon$, which limits the available statistics.

\begin{figure}
  \centering
  \includegraphics[width=\linewidth]{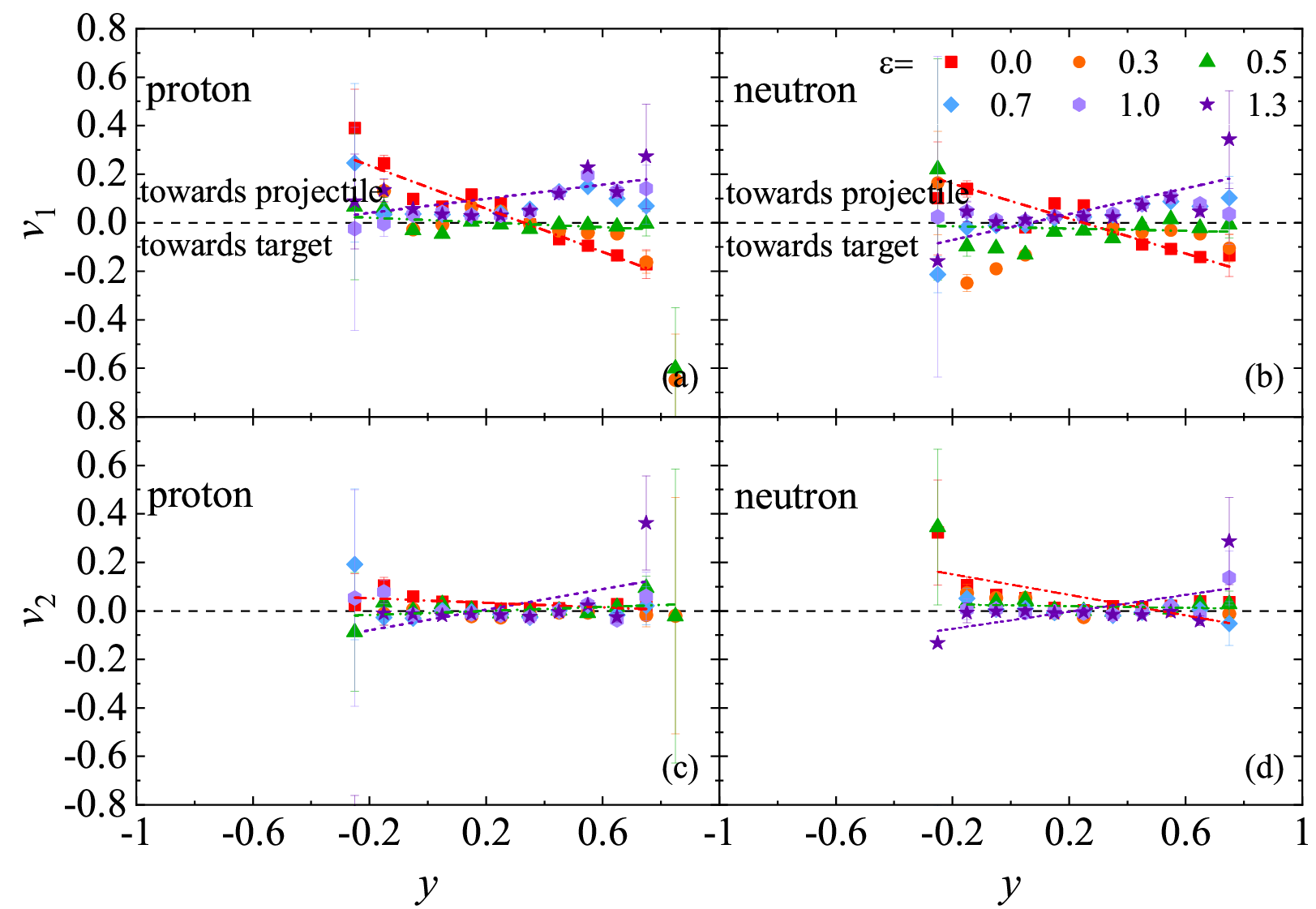}
  \caption{Directed flow $v_1$ and elliptic flow $v_2$ as functions of rapidity $y$ for protons (left) and neutrons (right). 
  The linear fits to data points of the corresponding colors reveal a systematic sign reversal in the slopes of both $v_1$ and $v_2$, transitioning from negative to positive as the isoscalar monopole excitation strength increases.
  }
  \label{fig:1H_1n_v12_y}
\end{figure}

Figure~\ref{fig:1H_1n_v12_y} shows the variations of $v_1$ and $v_2$ with rapidity $y$. The linear fits for $\varepsilon = 0.0$ (red dashed line), $\varepsilon = 0.5$ (green dash-dotted line), and $\varepsilon = 1.3$ (purple dotted line) demonstrate a systematic evolution: at low $\varepsilon$, the slope $dv_1/dy$ is negative, indicating net backward flow relative to the projectile frame, whereas at high $\varepsilon$ the slope becomes positive, signaling forward-peaked emission. Crucially, the same sign reversal occurs in the elliptic-flow gradient $dv_2/dy$, which transitions from slightly negative at low $\varepsilon$ to clearly positive at high $\varepsilon$. This parallel behavior demonstrates that monopole excitation does not merely modulate individual flow components but produces a systematic modification of the rapidity-dependent phase-space emission pattern. This behavior is consistent with the radial expansion of the excited $^{40}\mathrm{Ca}$ nucleus, which reduces binding energy and the effective surface thickness, thereby promoting early-stage forward emission of nucleons, especially at forward rapidities. As the system expands more vigorously at higher $\varepsilon$, the resulting anisotropic pressure gradients shift the directed-flow component $v_1$ toward the projectile side and modify the out-of-plane anisotropy characterized by $v_2$. These coordinated changes in $v_1$ and $v_2$ slopes may provide useful multidimensional signatures for identifying bubble-like density depletion even in asymmetric reactions.

\begin{figure}
  \centering
  \includegraphics[width=\linewidth]{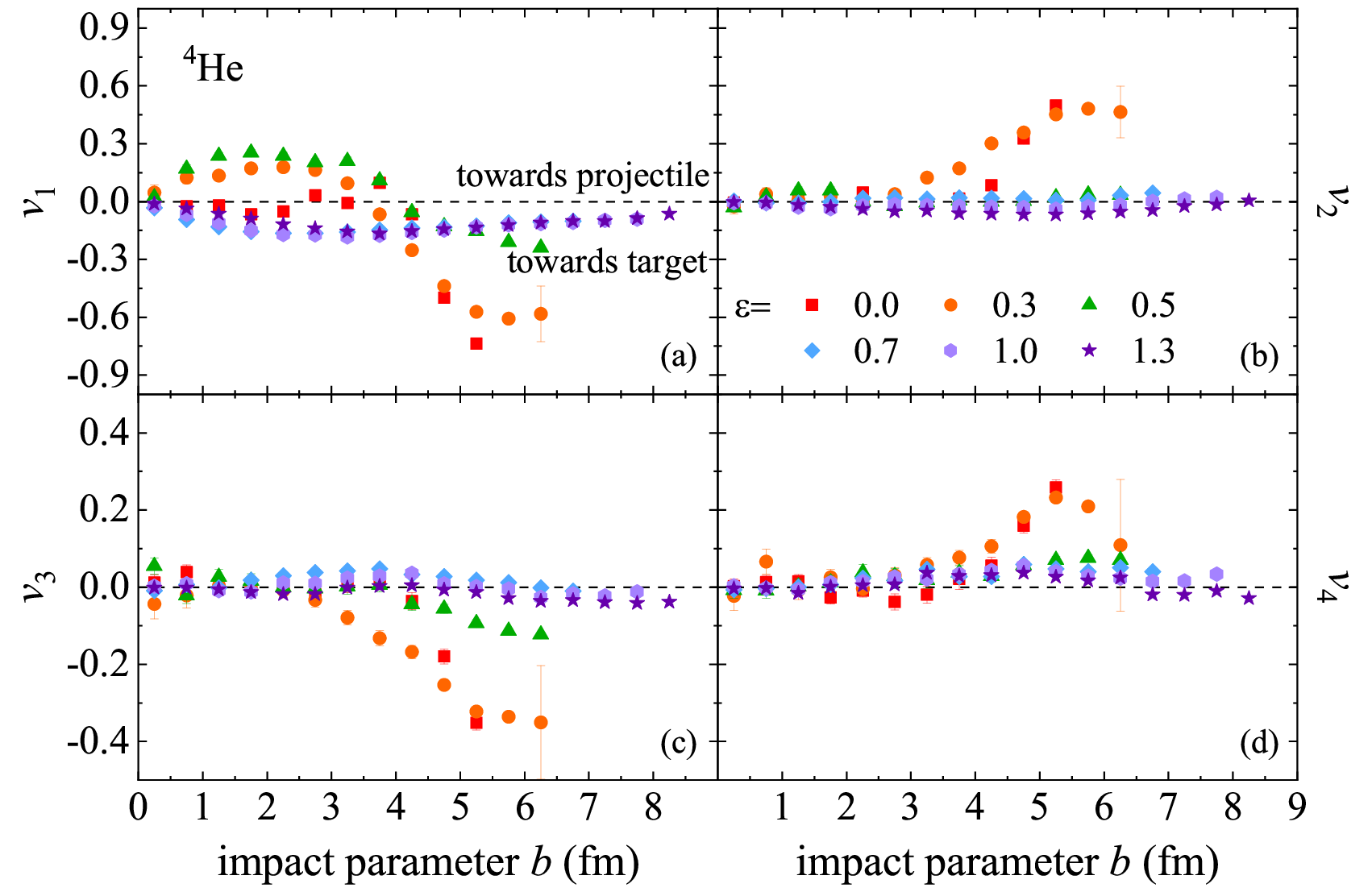}
  \caption{$v_n$-$b$ distributions of $^{4}\mathrm{He}$ under different $\varepsilon$. Panels (a)--(d) correspond to $v_1$--$v_4$, respectively. $^{4}\mathrm{He}$ fragments exhibit flow patterns distinct from those of protons and neutrons in Fig.~\ref{fig:1H_1n_v12_b} , especially in both sides of $\varepsilon\approx0.5$. }
  \label{fig:4He_v1234}
\end{figure}

The collective-flow response of $^{4}\mathrm{He}$ is also an important observable. In the relaxed low-energy cluster configurations of $^{40}\mathrm{Ca}$ and $^{9}\mathrm{Be}$, $\alpha$ particles ($^{4}\mathrm{He}$) are not merely fragmentation products but preformed structural units embedded in the initial cluster configuration, making $\alpha$ emission a dominant channel in this asymmetric system (Fig.~\ref{fig:CS}). Figure~\ref{fig:4He_v1234} shows that the response of the $\alpha$-particle flow coefficients $v_n$ to $\varepsilon$ differs from that of protons and neutrons. At low excitation, $v_1$ is positive in central to semi-central collisions but becomes negative at larger impact parameters. In the same peripheral region, $v_2$ and $v_4$ become positive, whereas $v_3$ becomes negative. For $\varepsilon>0.5$, all $v_n$ are close to zero, with $v_1$ and $v_3$ remaining slightly negative. Thus, stronger monopole excitation does not drive the $\alpha$-particle $v_1$ toward the projectile side but rather weakens the directional anisotropy of emission. This behavior is consistent with enhanced radial expansion and reduced cluster binding, which diminish coherent sideward deflection. The contrasting $\varepsilon$ dependence of $\alpha$ particles and single nucleons reflects their different dynamical responses and may provide a sensitive probe of cluster-driven dynamics in bubble-like configurations.

\begin{figure}
  \centering
  \includegraphics[width=\linewidth]{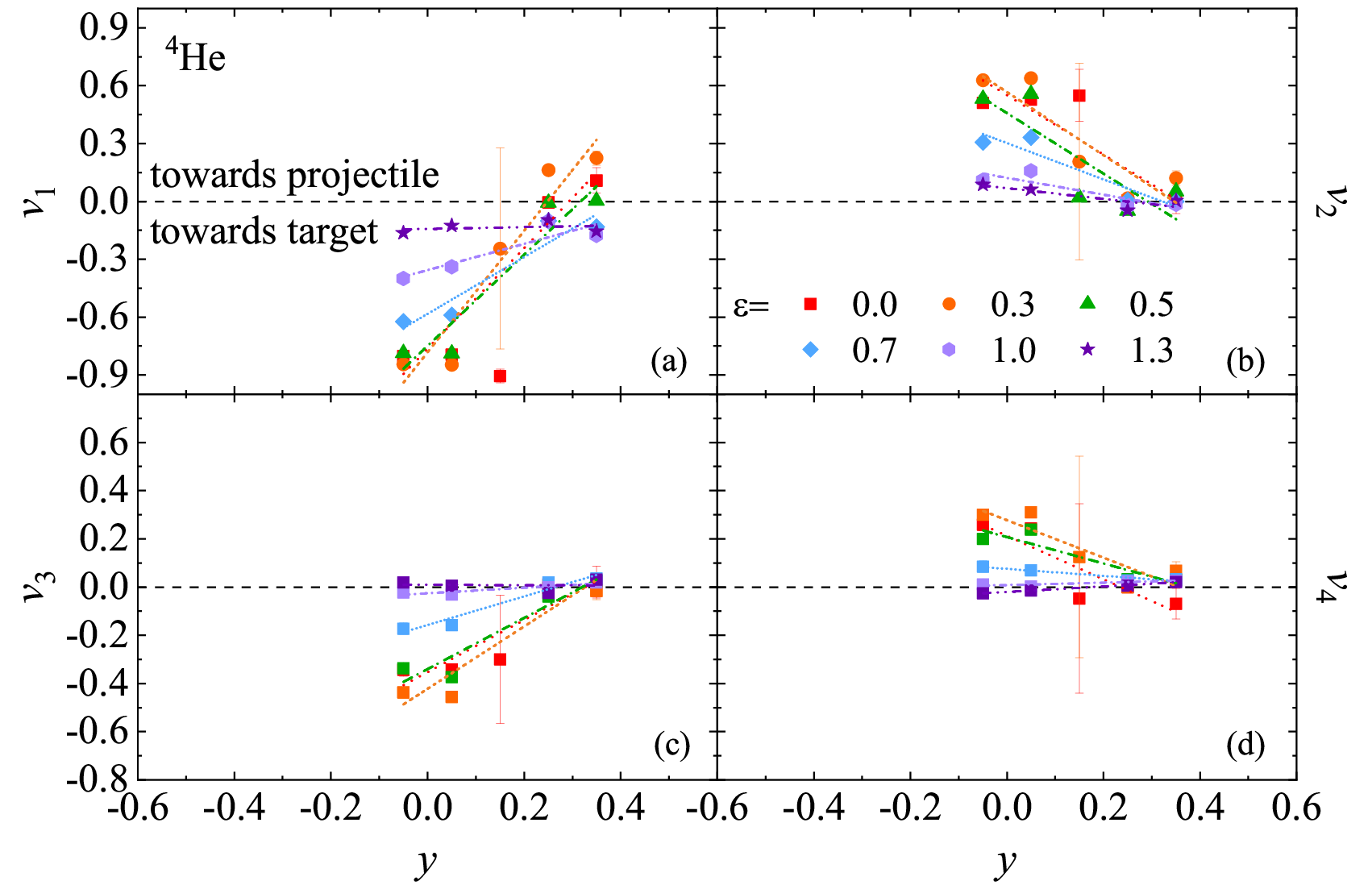}
  \caption{$v_n$-$y$ distributions of $^{4}\mathrm{He}$ under different $\varepsilon$. Panels (a)--(d) correspond to $v_1$--$v_4$, respectively. Lines denote linear fits to the data points with the corresponding colors.}
  \label{fig:4He_v1234_y}
\end{figure}

Figure~\ref{fig:4He_v1234_y} shows the $v_n$-$y$ distributions of $^4\mathrm{He}$ fragments, exhibiting a complementary evolution to that in Fig.~\ref{fig:1H_1n_v12_y} for protons and neutrons. As the monopole excitation strength $\varepsilon$ increases, the slopes $dv_n/dy$ systematically decrease and approach zero, particularly for $\varepsilon > 0.5$. Consistent with Fig.~\ref{fig:4He_v1234}, this behavior arises from the weakening pressure gradients acting on preformed $^4\mathrm{He}$ clusters as the bubble-like density structure expands, thereby suppressing directional emission anisotropy. Such evolution serves as a sensitive dynamical probe of cluster-configuration dynamics in nuclei with bubble-like low-density regions.

\section{Summary}
This study investigates collective-flow dynamics in the $^{40}\mathrm{Ca}+^{9}\mathrm{Be}$ projectile fragmentation reaction at 140~MeV/u, focusing on the influence of different isoscalar monopole excitation strength on fragment production and collective motion. Production cross sections, together with the collective-flow harmonics $v_1$--$v_4$ of protons, neutrons, and preformed $^{4}\mathrm{He}$ clusters, are systematically analyzed.

The calculations consistently reveal a distinct change in the nuclear response around $\varepsilon\approx0.5$, in agreement with the previously reported evolution of the density distribution and monopole excitation. Below this excitation strength, the reaction is dominated by the production of heavy fragments, while the directed flow $v_1$ of protons and neutrons remains predominantly negative, indicating preferential emission toward the target side. Above $\varepsilon\approx0.5$, the expansion of the low-density central region becomes significantly enhanced, leading to the suppression of heavy residues and enhanced production of light fragments. Simultaneously, both the magnitude and the rapidity dependence of the collective-flow observables exhibit pronounced changes, with the slopes of $v_1$ and $v_2$ reversing sign.

A distinct response is also observed for preformed $^{4}\mathrm{He}$ clusters. Owing to the initial ten-$\alpha$ cluster configuration adopted in the EQMD model, the emitted $\alpha$ particles represent preformed structural units rather than clusters formed solely during the reaction. Their collective flow differs from that of protons and neutrons, reflecting the distinct dynamical response of the cluster degree of freedom to monopole excitation. As the excitation strength increases beyond $\varepsilon\approx0.5$, the rapidity slopes $dv_n/dy$ of $^{4}\mathrm{He}$ fragments systematically approach zero, indicating a weakened directional anisotropy associated with the expansion of the bubble-like low-density region.

These results demonstrate that fragment production and collective flow provide complementary observables for probing the evolution of bubble-like nuclear density distributions. In particular, the correlated changes observed in density evolution, monopole response, production cross sections, and collective flow of protons, neutrons, and $\alpha$ particles around $epsilon \approx0.5$ suggest that this excitation strength marks the transition between two distinct dynamical response regimes of the bubble nucleus with different dencity distribution. The present results extend the study of bubble-like nuclear from static configurations and cluster dynamics monopole excitation to intermediate-energy nuclear reactions, and provides a new dynamical approach for identifying and distinguishing of bubble-like nuclear structures with different density distribution and their cluster degrees of freedom, thereby establishing dynamical probes for assessing the existenceof different size bubble nuclei.

\section{Acknowledgments}
This work is supported by the National Natural Science Foundation of China (NNSFC) (Nos. 12475134, 12375123, and U1832129), the Natural Science Foundation of Henan Province (No. 242300421048), the National Key Research and Development Program of China (Grant No. 2022YFA1602404), and the Strategic Priority Research Program of Chinese Academy of Sciences (No.~XDB34030000).

\bibliography{reference}

\end{document}